\documentclass{PoS}
\usepackage{subfigure}

\title{Detecting particles with cell phones: the Distributed Electronic Cosmic-ray Observatory }

\ShortTitle{The Distributed Electronic Cosmic-ray Observatory}

\author{\speaker{Justin Vandenbroucke}, Silvia Bravo, Peter Karn, Matthew Meehan, Matthew Plewa, Tyler Ruggles, David Schultz \\
Department of Physics and Wisconsin IceCube Particle Astrophysics Center, University of Wisconsin, Madison, WI 53706, USA\\
       E-mail: \email{justin.vandenbroucke@wisc.edu}}

\author{Jeffrey Peacock \\
Sensorcast
}

\author{Ariel Levi Simons \\
Loyola Marymount University
}

\abstract{
In 2014 the number of active cell phones worldwide for the first time surpassed the number of humans. Cell phone camera quality and onboard processing power (both CPU and GPU) continue to improve rapidly. In addition to their primary purpose of detecting photons, camera image sensors on cell phones and other ubiquitous devices such as tablets, laptops and digital cameras can detect ionizing radiation produced by cosmic rays and radioactive decays. While cosmic rays have long been understood and characterized as a nuisance in astronomical cameras, they can also be identified as a signal in idle camera image sensors. We present the Distributed Electronic Cosmic-ray Observatory (DECO), a platform for outreach and education as well as for citizen science. Consisting of an app and associated database and web site, DECO harnesses the power of distributed camera image sensors for cosmic-ray detection.
}

\FullConference{The 34th International Cosmic Ray Conference,\\
		30 July- 6 August, 2015\\
		The Hague, The Netherlands}

\begin{document}

\section{The DECO app}

\begin{figure}[htbp]
\begin{centering}
\includegraphics[width=0.4\textwidth]{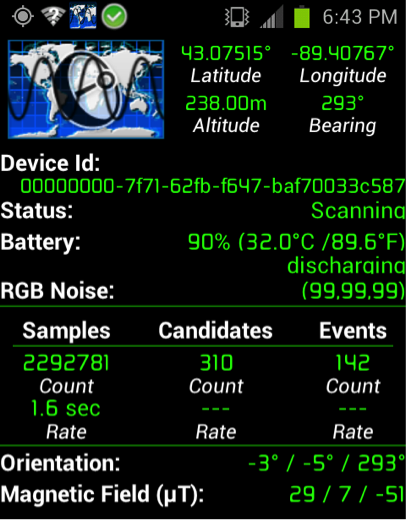}
\caption{Screenshot of the DECO app.  \label{deco}}
\end{centering}
\end{figure}

The DECO app (Figure~\ref{deco}) was released in September 2014 and can be downloaded at~\cite{decoWeb}.  It is currently supported on Android phones running version 2.1 or greater.  The app works by taking images with the camera sensor in repeated long exposures.  Each image is then analyzed to determine if it is a particle event candidate.  For efficient image processing, this occurs in two stages.  The first stage is a fast analysis of a low-resolution version of the image.  If the image passes this first filter, it is designated a ``candidate'' and reaches the second stage, which consists of the same analysis applied to a higher-resolution version of the image.  Each filter determines whether the image has a certain number of pixels above a certain threshold.  For each pixel the analysis is performed on the sum of the red, green, and blue color values.  Images passing both filters are designated ``events''.  In addition to the minimum pixel multiplicity requirement, a maximum pixel multiplicity requirement is also enforced in order to ignore images in which light is not sufficiently blocked from entering the camera.

Images are saved (with original red, green, and blue color values) with high quality (100\%) JPEG compression and then packaged along with metadata in a serialized format.  All images that pass both filter levels are retained.  In addition, every few minutes an image is saved regardless of whether it passed the filters.  This ``minimum bias'' data stream is useful for understanding variations in noise characteristics between individual devices (and between individual models).

A companion app, the data logger (downloadable at the same address) is used to synchronize data to a central database.  Each event has associated metadata including a unique device ID, geolocation, UTC time, altitude, phone orientation, magnetic field values, barometric pressure, temperature, model name, and Android version.  Images and metadata are automatically synchronized to the database every few minutes when a wifi connection is available.  If data are taken without an internet connection, data are automatically synchronized once one is available.  For best performance, the app should run with the phone plugged in to a power source and with the camera lens covered with electrical tape (to remove stray light that is a background for particle detection) and placed with the camera facing down on a flat surface.

\begin{figure}[htbp]
\begin{centering}
\includegraphics[width=0.8\textwidth]{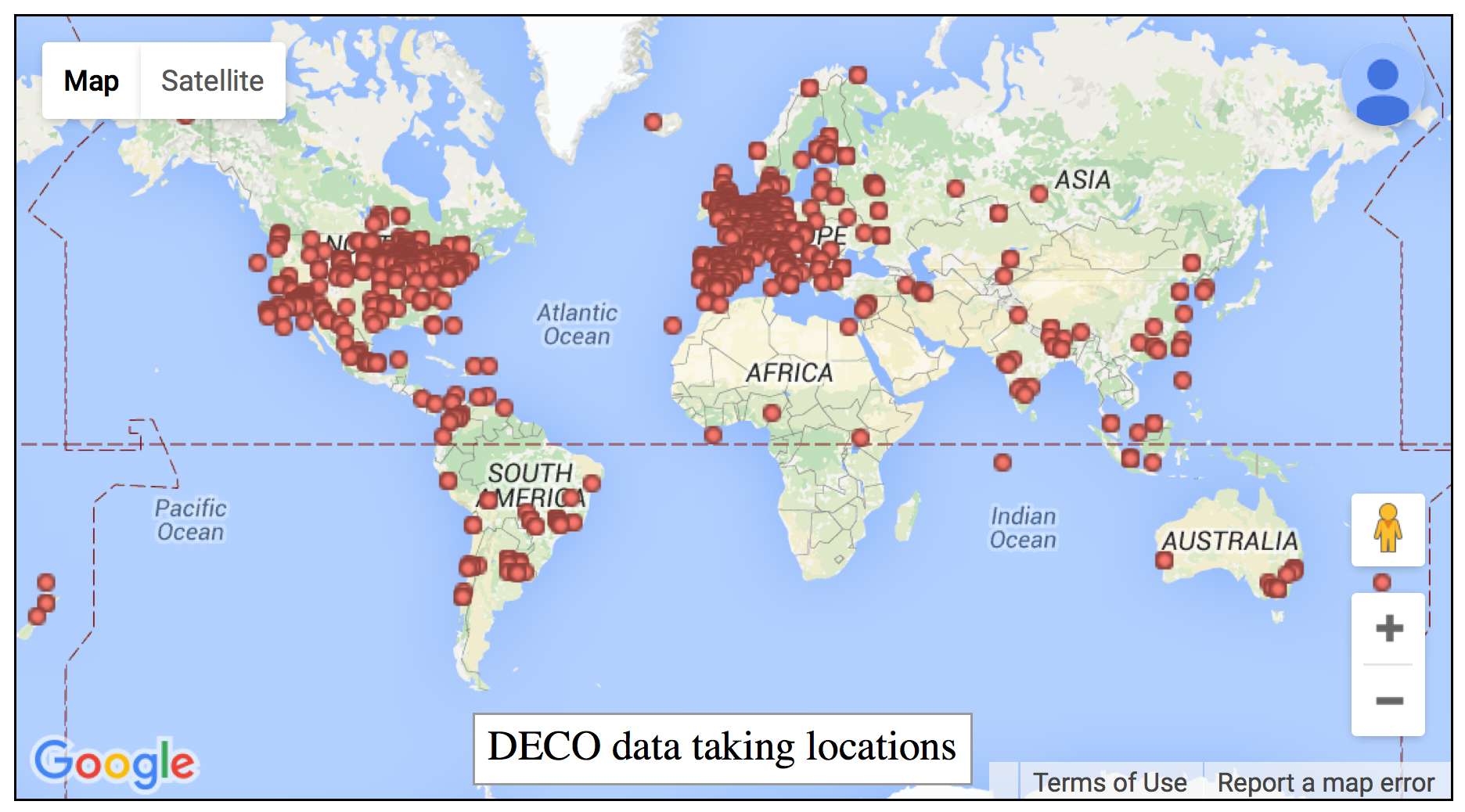}
\caption{Map of DECO data collection locations.  Data have been collected on all seven continents.  \label{map}}
\end{centering}
\end{figure}

The DECO app has been downloaded to thousands of devices of hundreds of different Android models.  The distribution of data taking locations is shown in Figure~\ref{map}.  DECO has taken data on every continent (including by scientists at the South Pole).

\section{Events and event types}

\begin{figure*}
\begin{center}
\subfigure[][]{
\label{a}
\noindent\includegraphics[width = 0.43 \textwidth]{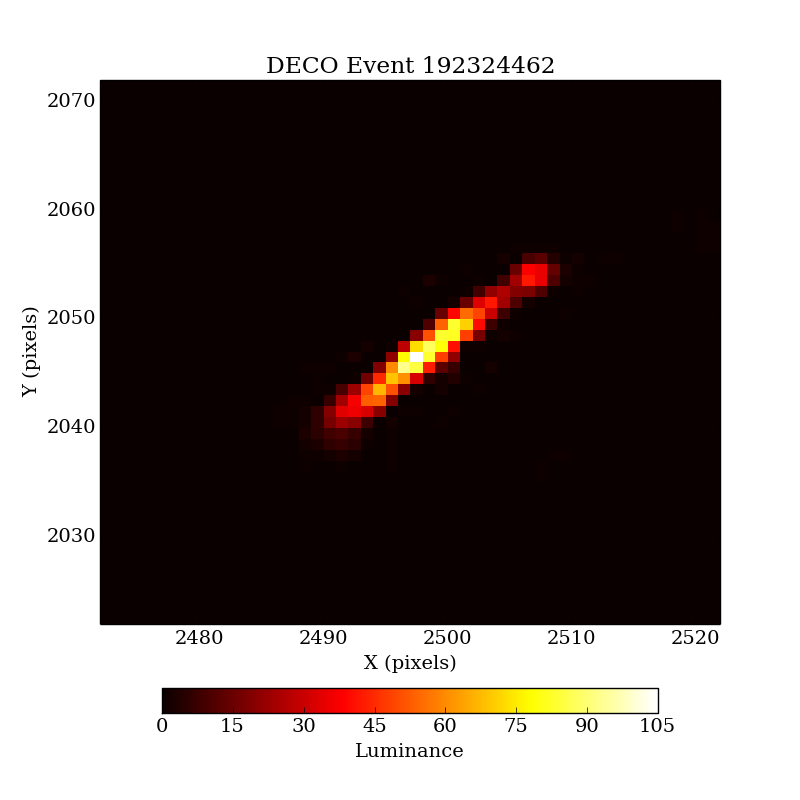}
}
\subfigure[][]{
\label{a}
\noindent\includegraphics[width = 0.43 \textwidth]{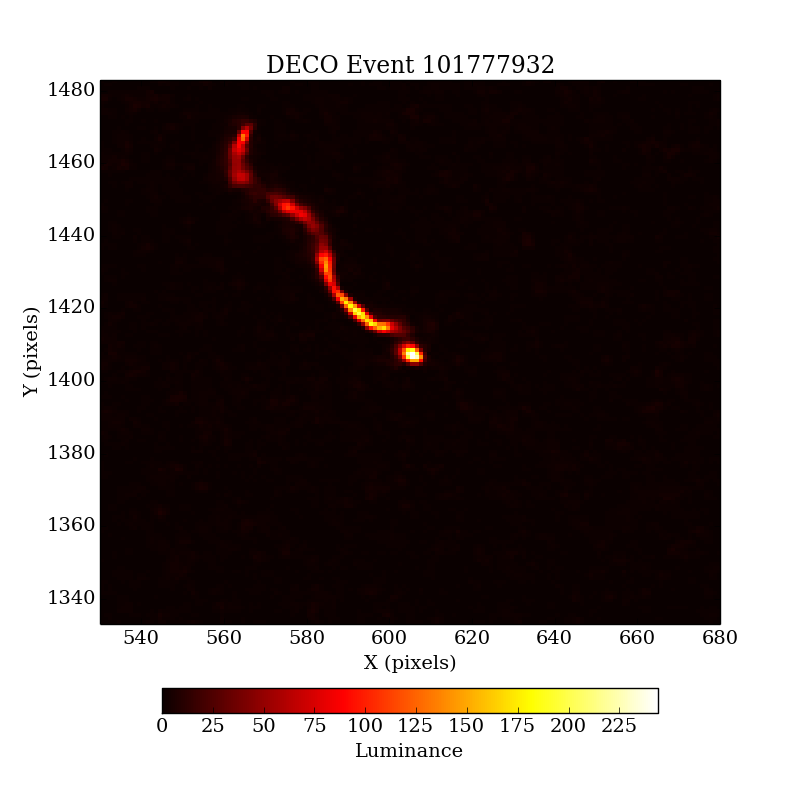}
}
\subfigure[][]{
\label{a}
\noindent\includegraphics[width = 0.43 \textwidth]{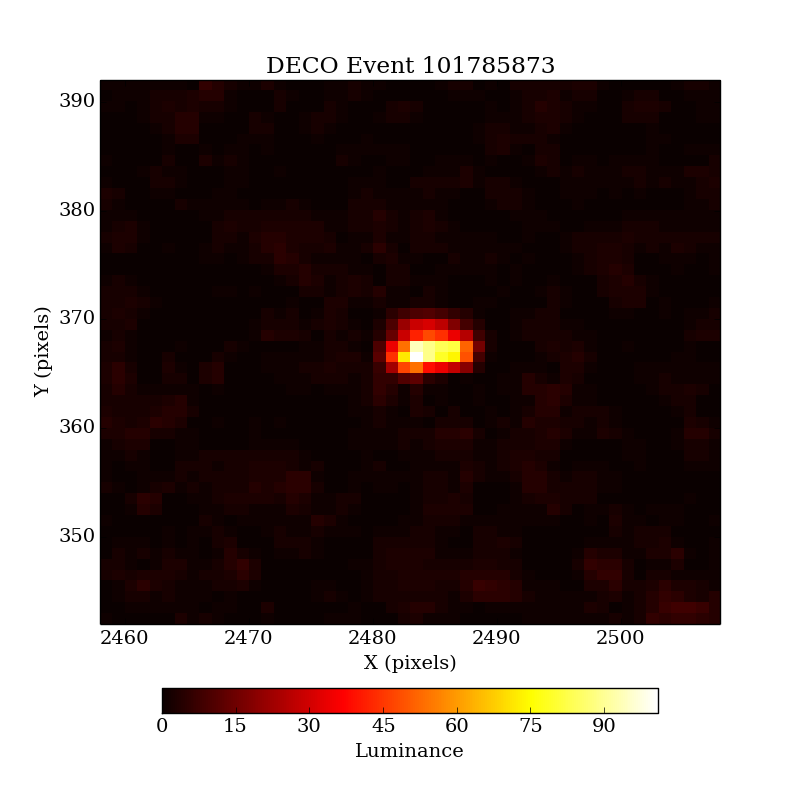}
}
\subfigure[][]{
\label{a}
\noindent\includegraphics[width = 0.43 \textwidth]{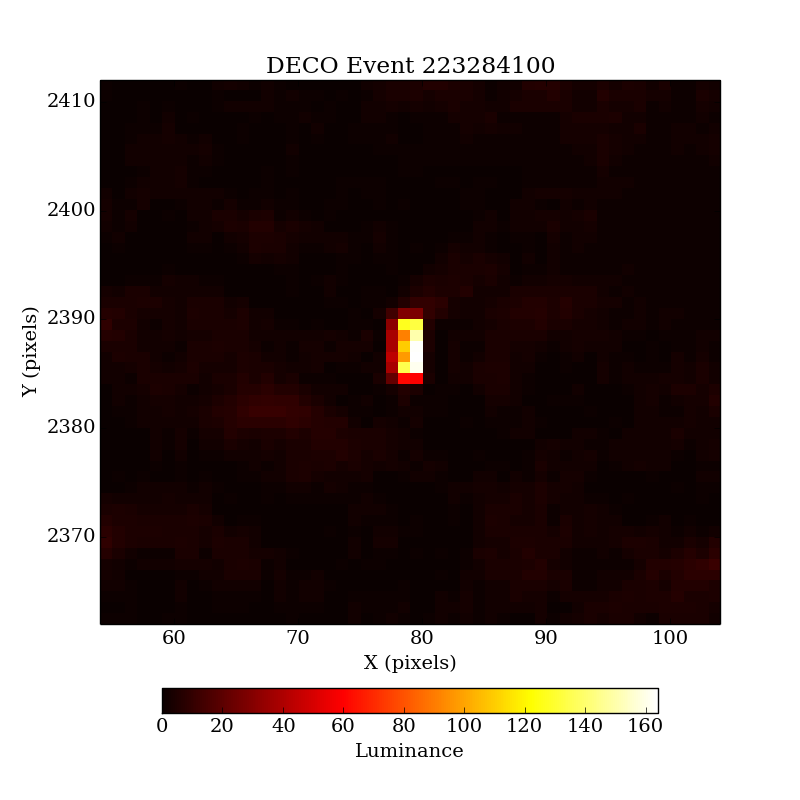}
}
\subfigure[][]{
\label{a}
\noindent\includegraphics[width = 0.43 \textwidth]{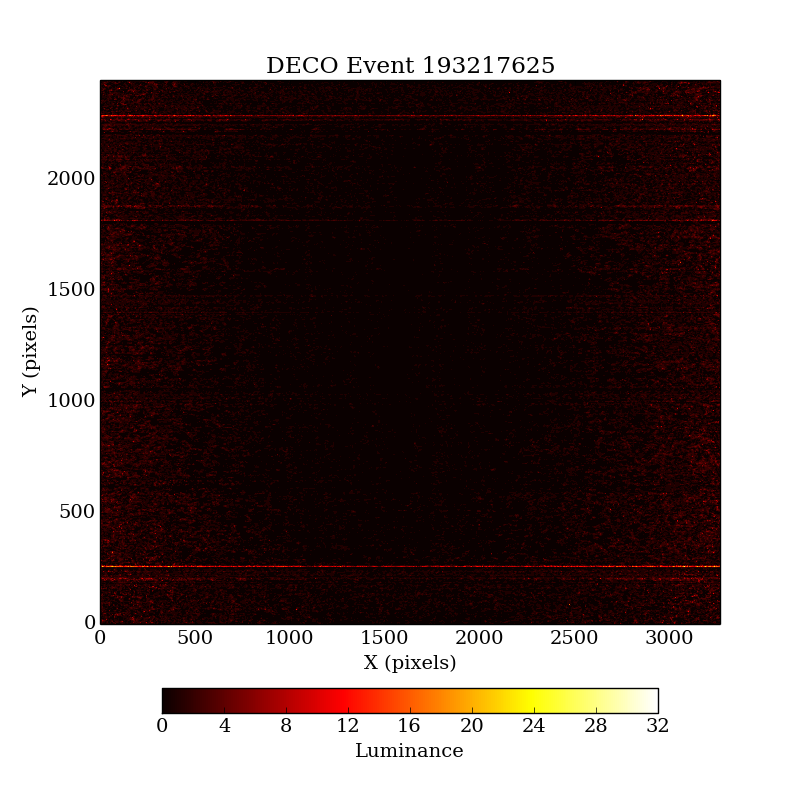}
}
\subfigure[][]{
\label{a}
\noindent\includegraphics[width = 0.43 \textwidth]{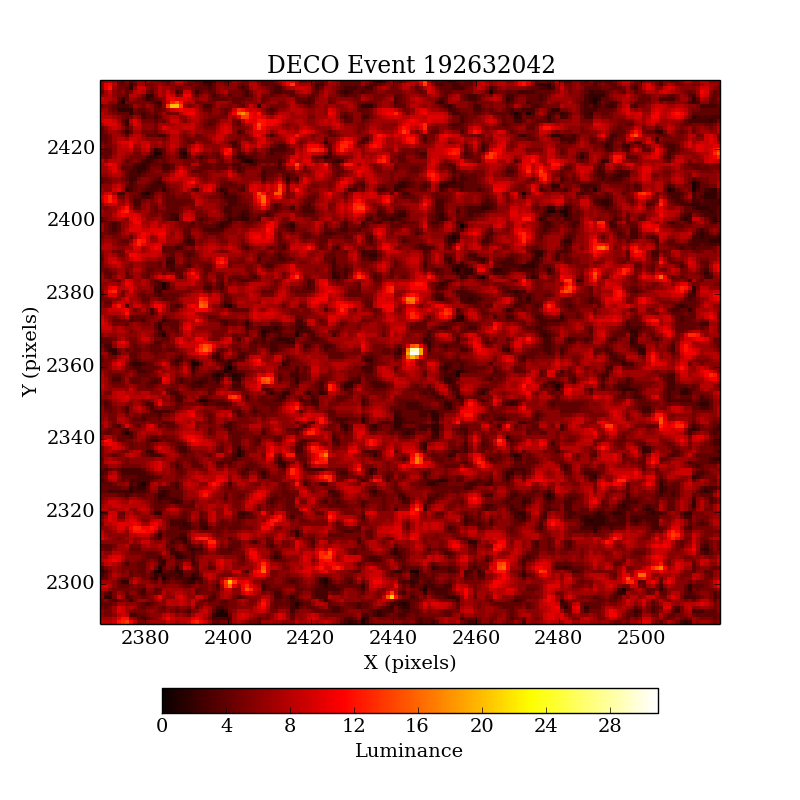}
}

\caption{Examples of DECO event classes: (a) track (b) worm (c) spot (d) hot spot (e) artifact (f) noise.  Luminance is a weighted sum of red, green, and blue color values.  The seventh event class, not shown, occurs when light is visible in the image.
\label{b}}
\label{events}
\end{center}
\end{figure*}

Figure~\ref{events} shows example images representing the event classes that DECO detects.  Particle event classes include those identified by~\cite{Groom} as \emph{tracks}, \emph{worms}, and \emph{spots}.  Tracks are long and straight, most often caused by high-energy (minimum ionizing, $\sim$GeV) cosmic-ray particles.  These can be secondary particles (especially muons) at sea level or primary particles (especially protons) at high altitude such as commercial airline cruising altitude ($\sim$30~kft).  Worms are due to low-energy ($\sim$MeV) electrons produced by radioactive decays in the phone material or surrounding environment.  The meandering worm trajectory is due to multiple Coulomb scattering of the electron.  In addition to the characteristic curved or wiggly shape within the sensor due to this scattering, the electrons can exit and re-enter the active (depleted) region of the sensor, resulting in multiple regions of energy deposition.  These worm topologies can either be produced by an incident beta particle or (more likely) by an incident gamma ray that Compton scatters to produce an energetic electron.  Finally, as argued by~\cite{Groom} and supported by data taken in undergound vs. surface and dirty vs. radio-clean locations, spots are likely due to lower-energy incident gamma rays that undergo a Compton scatter to produce a low-energy electron that is quickly absorbed.  A fraction of the spots could also be produced by cosmic-ray tracks traveling perpendicular to the sensor, or by alpha particles which have a short ($\sim$10~$\mu$m) range in silicon.  In some cases the tracks can produce delta rays, complicating their interpretation (especially considering that the recorded image is a relatively thin slice through a 3D set of particle energy depositions).  There are other events in which two or more tracks together form vertices.

In addition to the three categories of particle events, we identify four categories of sensor artifacts in the DECO events.  The first is simply caused by light reaching the image.  Due to the maximum pixel multiplicity requirement in the online filter, images that survive with light in them have a small amount of light.  The second category consists of thermal noise fluctuations.  Each image sensor exhibits noise which varies between device models and increases with temperature.  Some devices have high noise and upward fluctuations of the noise can cause an image to survive the online filter.  The third category is hot spots.  These occur when one or several neighboring pixels are consistently bright across multiple images.  They are clearly identified through their occurrence in a fixed position and shape between images.  The hot spots are smaller than the particle-induced spots.  Hot spots typically are not constant in brightness or even existence for a particular device, perhaps due to variation in thermal conditions near the image sensor.  Furthermore, for some device models the hot spots have been detected in the same position for different devices of the same model, which could be related to the device architecture and nearby heat sources in the electronics layout.  The fourth category is sensor artifacts in which multiple rows of pixels are bright.  We note that although the number of camera image sensor models is smaller than the number of Android device models, the variation in the Android hardware ecosystem is a challenge in interpreting the data.

In many event images of both the track and worm class, there is bright emission near one end of the particle trajectory.  This could be due to a Bragg peak.  This hypothesis is disfavored for GeV protons and muons which have a long range and are unlikely to stop within the sensor active region.  The Bragg peak interpretation is reasonable for MeV electrons.  An alternative hypothesis is that the brighter spot on one end (and not the other) is due to front-back asymmetry in the response of the sensor's depletion region to ionizing radiation.  If this hypothesis is correct, the effect could be used to break the forward-backward degeneracy in direction reconstruction.

\section{Database}

DECO data are automatically synchronized to a central database.  The data can be browsed and queried through a web-based interface available at~\cite{decoDataWeb}.  Users can view the events produced by their own devices or by others.   Although data are quickly (within a few minutes) synchronized from the devices, daily collating and processing of the data is necessary to organize it for the web interface.  Users taking data can therefore expect to view the data on the website within 24 hours of the acquisition.

Using the table of events available on the database web site, users can quickly view the images (zoomed and false colored versions similar to those in Figure~\ref{events}) as well as the geographic location of an event along with a link to the location in Google Maps.  Latitude and longitude are rounded to the nearest 0.01$^\circ$ for privacy.  Furthermore, users can query the data by selecting ranges of timestamps, latitude, longitude, altitude, device model, and device ID.  The web site displays a table of the first 50 events passing the selection, for fast online analysis.  This table shows the event timestamp, altitude, latitude, longitude, device model, and event ID.  Users can also download a .csv file with metadata for all events passing their  selection criteria.  The .csv file includes the quantities listed in the table in addition to others such as magnetic field components.  Students and interested members of the public can carry out simple or advanced analysis of the data using this web site.

\section{Depletion thickness measurement and direction reconstruction}

The CMOS sensors used for cameras feature pixels $\sim$1~$\mu$m square and depletion region thickness between 10~$\mu$m and 100~$\mu$m.  This geometry combined with the angular distribution of the incident particles determines the track length distribution detected by DECO.  We can therefore combine the track length distribution with the known cosmic-ray angular distribution at sea level to measure the depletion thickness of the sensor.  While the pixel width is often publicly available, the depletion thickness is not.

Combining the angular distribution of cosmic rays ($\cos^2\theta$, where $\theta$ is the zenith angle), the sensor effective area as a function of zenith angle, and the solid angle as a function of zenith angle, we derive the distribution of track lengths $L$ in terms of the depletion thickness $H$ (both measured in units of the pixel width).

$$\frac{dN}{dL} \propto \frac{L H^4}{(L^2 + H^2)^{3}} $$

\begin{figure}[htbp]
\begin{centering}
\includegraphics[width=0.6\textwidth]{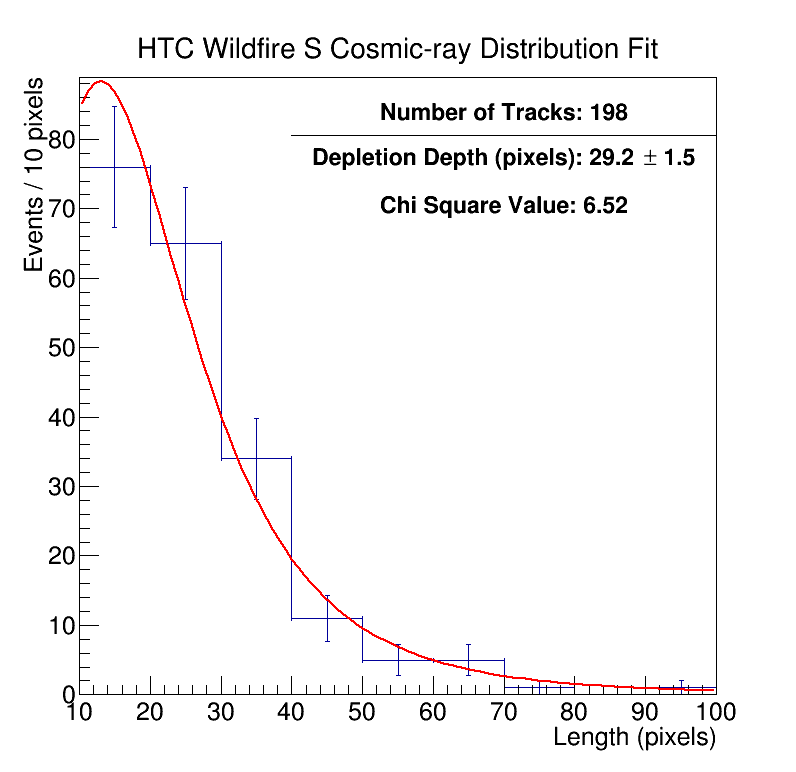}
\caption{Distribution of measured track lengths for good muon candidate events, along with a fit of the distribution expected from the angular distribution of cosmic rays at sea level.  The fit has a single free parameter corresponding to the thickness of the active (depleted) sensor volume.  The chi-squared value is $6.52$ for $9-1=8$ degrees of freedom.\label{depletion}}
\end{centering}
\end{figure}

We analyzed a particular device model, the HTC Wildfire S.  We first applied image classification cuts to select track-like events and reject spots and worms.  We required a minimum track area in order to reject spots, and a minimum track eccentricity to reject worms.  The track length distribution after these cuts is shown, along with a fit of the above derived distribution, in Figure~\ref{depletion}.  The fitted depletion thickness is 29.2 $\pm$ 1.5 pixels.

According to ~\cite{wildfire}, this camera image sensor is 2400~$\mu$m $\times$ 1800~$\mu$m for its 2592 $\times$ 1944 pixels, corresponding to a 0.9~$\mu$m square pixel size.  This means that the DECO measured depletion thickness is 26~$\mu$m.  This depletion thickness can be used to determine the zenith angle of individual muons producing tracks.  Combined with the absolute orientation of the device, this means that the original direction from which the muon arrived can be determined, up to forward-backward degeneracy along the track.  This degeneracy may be resolvable in the future using front-back differences in the response of the active sensor region as discussed above.

\section{Conclusion and outlook}

The DECO project provides a cell phone app and interactive web site for education and outreach as well as citizen science.  Students and other interested members of the public can contribute to both data collection and data analysis.  Algorithms are under development for automatic classification of events into the seven categories described above.  While there are some background events due to camera artifacts and noise, signal events from both cosmic rays and ambient radioactivity are clearly detected.

We note that because the phone orientation is logged in the data stream, event reconstruction can be used to determine the incident direction of individual cosmic rays.  Relative to the device, the azimuthal direction can be determined from the track direction in the image plane, and the polar direction (zenith angle, for a horizontal phone) can be estimated with less precision using the depletion thickness of the sensor combined with the length of the track.  Finally, it may be possible to break the forward-backward ambiguity along the track direction (e.g., to distinguish east-going from west-going) by using asymmetry in the front-back response of the sensor.

For use in the classroom and at home, the DECO project enables people to use devices they already carry in their pockets as particle detectors, making it easy for them to engage through hands-on data taking and  analysis in the science of cosmic rays and radioactivity.  The app and database can be used either for short measurements and data browsing or for longer data taking and analysis projects such as comparing rate vs. altitude in different cities, measuring rate vs. altitude using airplane and balloon flights, and potentially measuring the East-West effect through which cosmic rays were first determined to have positive charge.  The ubiquity of cell phones and other mobile devices makes DECO more accessible to a wider range of schools and audiences than projects that use more sophisticated instrumentation such as scintillators.  DECO therefore provides a complementary approach to scintillators, which have greater sensitivity due to their much larger effective area but require more setup and expertise, are more expensive, and are not as widely available.

While the power of DECO for outreach and education is clear, the long-term prospects for making new scientific measurements are not yet known.  It is possible that cell phones could be used to make new discoveries by correlating cosmic-ray rates sampled at many locations and times with other data sets, or through detection of extensive air showers~\cite{crayfis}.  The challenges to achieving this are significant: inhomogeneity of the hardware and data acquisition conditions, discrimination between cosmic-ray tracks and low-energy radioactivity events, limitations in multi-device coincidence set by the poor time resolution determined by the exposure duration (milliseconds to seconds, rather than nanoseconds to microseconds as for other cosmic-ray detectors), and the extremely large number of participants that would be necessary to detect ultra-high-energy cosmic rays through multi-device coincidence~\cite{criticism}.  Finally, we note that increasingly advanced hardware, firmware, and software in cell phone cameras includes processing to remove noise, which can pose a challenge to particle detection.  There is evidence in the DECO data that some of the newest devices have significantly lower rates than older devices.

\section{Acknowledgments}

DECO is supported by the American Physical Society, the Knight Foundation, the Simon-Strauss Foundation, and QuarkNet.  We are grateful for beta testing, software development, and valuable conversations with Keith Bechtol, Segev BenZvi, Andy Biewer, Paul Brink, Patricia Burchat, Duncan Carlsmith, Alex Drlica-Wagner, Mike Duvernois, Lucy Fortson, Stefan Funk, Mandeep Gill, Laura Gladstone, Giorgio Gratta, Jim Haugen, Kenny Jensen, Kyle Jero, David Kirkby, David Saltzberg, Marcos Santander, and Ian Wisher.

\bibliography{references}
\bibliographystyle{elsart-num}


\end{document}